\documentclass[manuscript,screen]{acmart}
\AtBeginDocument{%
  \providecommand\BibTeX{{%
    \normalfont B\kern-0.5em{\scshape i\kern-0.25em b}\kern-0.8em\TeX}}}

\setcopyright{acmcopyright}
\copyrightyear{2022}
\acmYear{2022}
\acmDOI{XXXXXXX.XXXXXXX}
\usepackage{makecell}
\usepackage{multirow}
\usepackage{subcaption} 
\usepackage{multicol}
\usepackage{caption}
\newcolumntype{C}[1]{>{\centering\arraybackslash}m{#1}}
\usepackage{array}
\usepackage{adjustbox}
\usepackage{soul}

\acmConference[Conference acronym 'XX]{Make sure to enter the correct
  conference title from your rights confirmation emai}{June 03--05,
  2018}{Woodstock, NY}
\acmBooktitle{AMX '22: ACM Interactive Media Experiences
 June 22--24, 2022, Aveiro, Portugal} 
\acmPrice{15.00}
\acmISBN{978-1-4503-XXXX-X/18/06}

\begin{document}

\title{TV-watching partner robot: Analysis of User's Experience}

\author{Donghuo Zeng}
\authornote{Both authors contributed equally to this research.}
\email{do-zeng@kddi-research.jp}
\orcid{1234-5678-9012}
\author{Jianming Wu}
\authornotemark[1]
\email{ji-wu@kddi-research.jp}
\affiliation{%
  \institution{KDDI Research, Inc.}
  \streetaddress{2-1-15 Ohara, Fujimino, Saitama}
  \city{Saitama}
  \country{Japan}
  \postcode{356-8502}
}

\author{Gen Hattori}
\affiliation{%
  \institution{KDDI Research, Inc.}
  \streetaddress{2-1-15 Ohara, Fujimino, Saitama}
  \city{Saitama}
  \country{Japan}}
\email{ge-hattori@kddi-research.jp}
\author{Yasuhiro Takishima}
\affiliation{%
  \institution{KDDI Research, Inc.}
  \streetaddress{2-1-15 Ohara, Fujimino, Saitama}
  \city{Saitama}
  \country{Japan}}
\email{takishima@kddi-research.jp}

\renewcommand{\shortauthors}{Donghuo and Jianming, et al.}

\begin{abstract}
Watching TV not only provides news information but also gives an opportunity for different generations to communicate. With the proliferation of smartphones, PC, and the Internet, increase the opportunities for communication in front of the television is also likely to diminish. This has led to some problems further from face-to-face such as a lack of self-control and insufficient development of communication skills. This paper proposes a TV-watching companion robot with open-domain chat ability with a range of 50 daily life topics. The robot contains two modes: “TV-watching mode” and “conversation mode”. In “TV-watching mode”, the robot first extracts keywords from the TV program and then generates the disclosure utterances based on the extracted keywords as if enjoying the TV program. In the “conversation mode”, the robot generates question utterances with keywords in the same way and then employs a topics-based dialog management method consisting of multiple dialog engines for rich conversations related to the TV program.  We conduct the initial experiments and the result shows that all participants from the three groups enjoyed talking with the robot, and the question about their interests in the robot was rated 6.5/7-levels. 
This indicates that the proposed conversational features of TV-watching Companion Robot have the potential to make our daily lives more enjoyable. Under the analysis of the initial experiments, we achieve further experiments with more participants by dividing them into two groups: a control group without a robot and an intervention group with a robot. The results show that people prefer to talk to robots because the robot will bring more “enjoyable”, “relaxed”, “and interesting”.

\end{abstract}



\maketitle

\begin{figure}[htp]
  \includegraphics[width=\textwidth]{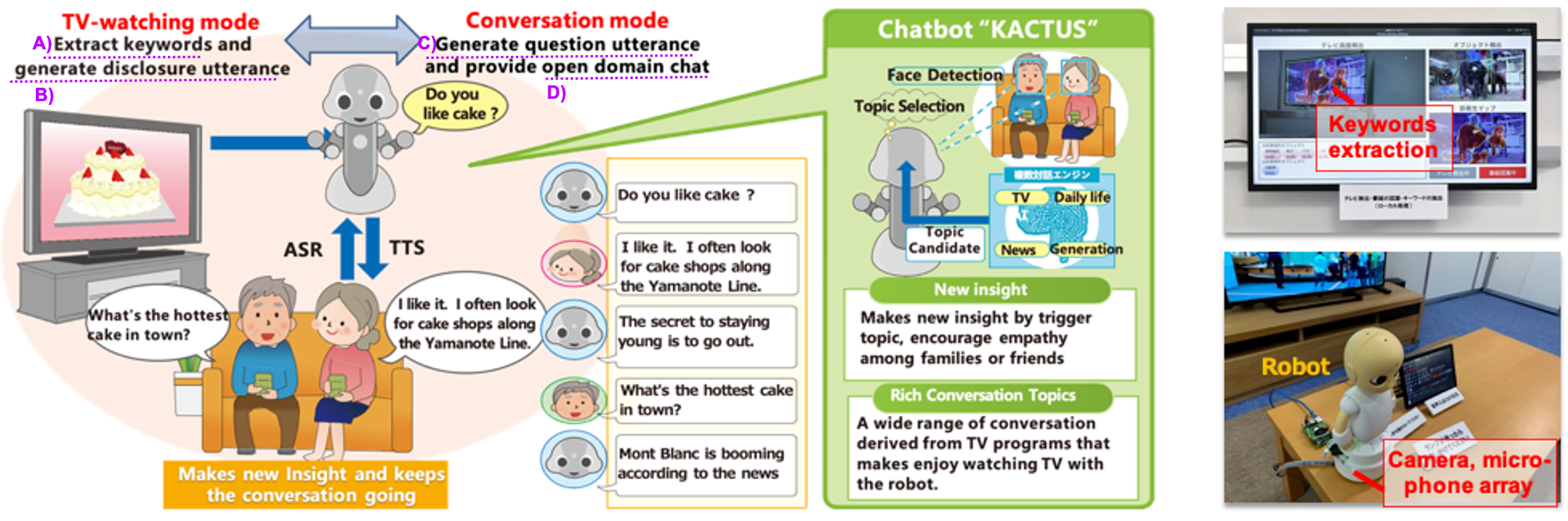}
  \caption{The concept of TV-watching companion robot.}
  \Description{Enjoying the baseball game from the third-base
  seats. Ichiro Suzuki preparing to bat.}
  \label{fig:teaser}
\end{figure}

\section{Introduction}
Televisions (TVs) have become a ubiquitous part of our daily life and most homes have one in their living room. Watching TV in the living room not only allows you to get news information, enjoy experience entertainment programs, and gain some knowledge regarding daily life, but also improves communication among family members from different generations or friends. In this case, the individuals have the opportunity to enter into conversations and feel joy or sorrow through topics related to the television programs they watch. On the other hand, with the proliferation of smartphones, personal computers and the Internet, it is now possible to obtain information from sources other than television, so there may be fewer opportunities to communicate in front of the television. This reduction in communication results in fewer opportunities to interact with others, leading to the problems we observe today, such as health risks~\cite{house1988social}, inadequate development of self-control and poor communication skills~\cite{sok001}. 

With the development of AI technology, social robots have become widespread in recent years. In the future, it is expected that they will become further involved in our lives. However, currently, robots are not able to fluently communicate with humans. One reason for this problem is that the robot cannot behave/respond based on context that is not obvious in a conversation, whereas people can have a conversation based on the context behind the conversation. When we watch TV with our family or friends, we usually enjoy watching TV by sparking conversations and sharing our emotions through the contents of the programs and topics discussed by others. Therefore, we believe that by using information from TV programs can be used as the shared context for the robot to communicate with people. In addition, it may be possible to activate the communication while watching TV. In particular, the conversational robot/chatbot is expected to help the elderly improve the effectiveness of their interactions and reduce their feelings of loneliness and isolation. This could slow the rate of disability in the elderly and reduce social costs and the burden on caregivers~\cite{su2017chatbot}

Since previous research~\cite{yamamoto2009dialogue} has demonstrated the acceptability of the conversational robot with a single person living alone at home, we would like to focus on the conversational features of the TV-watching robot that accompanies families or friends to accept the future user experience of watching TV with a robot. Therefore, our work is guided by the following two research questions. 

\begin{itemize}
\item {\verb|Research question 1:|} Will the user experience be acceptable when family or friends watch TV with the conversational robot?
\item {\verb|Research question 2:|} Do the robot's conversational features improve the well-being and relationships of families or friends?
\end{itemize}

In this case, we propose a TV-watching companion robot based on the conversational analysis of personal conversation behaviors when watching TV~\cite{hoshi001, hagio001}.
We utilize an open-domain chatbot "KACTUS"~\cite{kactus001, wu2021tv} on the robot and expect it to be create even more excitement when watching on a TV screen with friends, and have the effect of encouraging communication. Related work is presented in Section 2. The system design of our robot prototype is presented in Section 3, and Section 4 presents the results of the TV watching experiment using the robot prototype. The study is summarized in Section 5.

\section{Related works}
\subsection{TV robots}
Although virtual agents that interact with people can be in the form of either computer-generated(CG) displays, robots in real space are reportedly more suited to dialog concerning an object that actually exists than a virtual agent on a display~\cite{kidd001}. Therefore, there have been attempts to use robots in TV watching environments.

Ogawa et al. proposed a method for operating IoT devices, including robots, by distributing metadata in tandem with TV programs~\cite{ogawa001}. The advantage of this approach is that the robots can be operated operated in concert with the TV program being watched. However, the cost of producing TV programs for broadcasters could increase owing to the creation metadata for operating robots. Nishimura et al. proposed a robot that uses comments posted on social media such as Twitter to chat with users\cite{nishimura001}. This approach does not require the broadcaster to create metadata to be used for operating robots. However, when users are watching a TV program with low ratings, the number of robot utterances decreases because there are fewer comments on social media. Minami et al. proposed a TV watching chatting robot using social networking service (SNS) comments to incorporate humor so that users do not tire of robots and discontinue their use~\cite{minami001}. Muto et al. proposed a robot system architecture and algorithm providing TV programs and related information based on the user's interests~\cite{muto001}. However, none of these robot systems is based on an open domain chatbot.

On the other hand, HMD methods have been proposed that provide additional information to an existing TV program~\cite{ARTV001, ARTV002}. Robots in real space are reportedly better at conversing in real space than a virtual agent on a displays~\cite{kidd001}. However, the additional devices must be attached to the human head or body, which is not just a burden but also is disruptive to conversations between people.

\subsection{Personal conversation behaviors when watching TV}
To understand the structure of natural dialog between persons watching TV, we recorded and analyzed dialog when groups of two well-acquainted people freely watched their favorite programs~\cite{hoshi001}. We have conducted dialog analyses by classifying the types of conversations that people have among themselves when watching TV in order to determine their behavior. It was found that the percentage of "disclosure" utterances (when an individual states their feelings or thoughts) in watching TV and "question" in conversation was high. 
\subsection{Human-robot dialog with elderly people}
It is a matter of urgency to develop a human-robot dialog system for the elderly people given the high percentage of population who aged 65 or over, such as in Japan~\cite{Annualreport}, to help those who live alone increase their opportunities to have a conversation in daily life. The Pearl~\cite{pollack2002pearl} reminds elderly people to perform daily activities, such as going to the bathroom every three hours and taking medication. In addition, the caregiver of an elderly people can enter enter someone’s daily activities in advance, and the Pearl will reminds the person based on the schedule. The first trial system~\cite{minami2016chat} was designed to support a user to chat with a robot while watching a TV program, and it achieved natural timing in terms of simultaneous responses and supported interesting discourse derived from social media networks related to the TV program the user was watching. One of the barriers in human-robot dialog system development is that speech recognition frequently fails. To overcome the problem, a proposed question–answer–response dialog model~\cite{iio2020twin} enable the robot to actively asks the users various questions and makes it possible for two robots to participate in the dialog. Another study ~\cite{jokinen2019human} aimed at dialog modelling to enhance the communication capability between users and robots to improve the elderly people care services provided to elderly people. Furthermore, they introduced a dialog system architecture that has the ability to communicate with users based on care-giving tasks.

The conversational robots in the past conducted mostly directed or limited topic conversations, we propose a conversation robot for open domains that can freely talk about richer topics and behave like a human.

\section{SYSTEM DESIGN}
Our conversational features is designed as the following principles, which is shown in Fig.~\ref{fig:teaser}. the robot works by switching between the two following modes, "TV-watching mode" and "Conversation mode” to simulate human conversation behaviors when watching TV~\cite{hoshi001}. The robot prototype could be switched repeatedly between "TV-watching mode" and "conversation mode". 

\textbf{1) "TV-watching mode":} In this mode, the robot talks about the commentary and the impression it gets from the content of TV as if it is enjoying the TV program. We think that the robot might initiate a conversation topic to create communication opportunities between families or friends when they are watching TV. We implement the following two features as follows. 
\begin{itemize}
\item {\verb|Feature A:|}  Extracts keywords from the TV program
\item {\verb|Feature B:|}  Generates the disclosure utterances based on the extracted keyword
\end{itemize}

\textbf{2) "Conversation mode":} In this mode, the robot generates question utterances with keywords in the same way and then offers a chat in an open domain related to the TV program. The conversation topic start related to the TV program, then it will expand to daily life and the latest News/SNS. We think that the robot might give more communication opportunities to improve the well-being and relationships of families or friends. We implement the following two necessary functions as follows. 
\begin{itemize}
\item {\verb|Feature C:|} Generates question utterances with keywords from the TV program
\item {\verb|Feature D:|}  Employs a topics-based dialog management method consisting of multiple dialog engines for rich conversations related to the TV-program
\end{itemize}

Our robot prototype, equipped with the proposed open-domain conversational and an extension of the development robot CommU~\cite{commu001}, is shown in Fig.~\ref{fig:teaser}. We introduced a prototype of camera-microphone array and installed it as an additional device at the bottom of the CommU (Fig.~\ref{fig:teaser} bottom right).
%
Using in the camera-microphone array, the robot could detect people nearby in the camera-microphone array within a range of approximately 200 degrees and distinguish the difference between a human voice and TV sound by means of the HARK's sound localization and separation method~\cite{hark001}. The distinguished human voice was converted into text by Microsoft speech recognition~\cite{azure001}. Furthermore, although the study presents each function in English, the robot actually operated in Japanese.

\subsection{TV watching mode (including the feature A and B)}
Feature A: Firstly, the robot obtained the video, audio, and caption data from the TV program to extract keywords for the utterance generation. These sentences in the caption data were converted into word-separated 
writing using MeCab~\cite{mecab001} with mecab-ipadic-NEologd~\cite{mecab002}. If the TV programs do not contain caption data, we extract keywords from the content of TV program by implementing a video object detection method~\cite{rcnn001, coco001} on the AWS~\cite{aws001}, which learns by MS COCO~\cite{coco001r} and employs recognition API provided by ~\cite{azure001}. 

The classification results of the detected objects are used as the extracted keywords. An example of keyword extraction using object detection on a local computer is shown in Fig.~\ref{fig:teaser}. The green rectangle indicates the object detected and the red area indicates the high salient estimation results. In the example of figure, the “elephant” was extracted as the keyword. We used Microsoft Azure~\cite{azure001} and Amazon Web Service~\cite{azure001} for the cloud services~\cite{aws001}. The keyword extraction of “elephant” is shown in the upper right of Fig.~\ref{fig:teaser}), as an example.

Feature B: Then, the robot’s disclosure utterances which express the robot’s feelings, are generated using the extracted keywords. We used the captions of the TV programs on channel 10 over the past seven years as the template sentences corpus for the disclosure utterance generation. These template sentences have fewer than 20 characters, because 79\% of all utterances on TV program are less than 20 characters~\cite{hoshi001}. The template sentences of past captions were used to learn the distributed representation of words by the Word2Vec~\cite{vec001}. The robot selects the template sentence to use by calculating the high cosine similarity to the extracted keywords. Then, an utterance is generated by inserting/replacing the new keyword into the template sentence. As in the above example that the a disclosure utterance as “I like elephants” will be generated. Some template samples are as follows.
\begin{itemize}
\item Disclosure: The generated disclosure utterances used template sentences containing emotional expressions
(e.g., I want to eat ***, I want to go ***: keywords is expressed as *** )
\item Question: The generated disclosure utterances used template sentences containing questions (e.g., Do you want to eat ***, Do you like ***: keywords is expressed as *** )
\end{itemize}

Next, the robot determines whether to switch to the conversation mode. The proportion of disclosure and question speeches were determined in advance as parameters, and either of the two were randomly selected based on this proportion. The utterance interval could also be adjusted to avoid situations where the robot is constantly speaking. The utterance frequency was set in advance, and the utterance interval was determined using a Poisson distribution. Furthermore, keywords used once in utterance generation would not be reused for a while to generate a variety of utterances.

Finally, the robot turns to the TV and transits to the conversation mode to say the generated disclosure utterances by the CommU TTS API~\cite{commu001}. At the same time as the above processing, the robot facing the TV continuously repeats random operations such as "blinking", "nodding", "moving the neck", "moving the upper body", just acting as though it is watching TV.

\subsection{Conversation mode (including the feature C and D)}
Feature C: In this mode, the question utterances are generated in manner similar to the above-described disclosure utterances. We used the the same template sentences corpus for the generation of question utterances. The robot selects which template sentence to use by calculating the cosine similarity of the extracted keywords. As in above example in the Fig.~\ref{fig:teaser} that the question utterance as “Do you like elephant?”, will be generated.

The proportion of disclosure and question utterances was determined in advance as parameters, and either of the two were randomly selected based on this proportion. The utterance interval could also be adjusted to avoid situations where the robot is constantly speaking. Furthermore, keywords used once in utterance generation would not be reused for a while to generate a variety of utterances.

Feature D: At this time the robot turns to the viewers and says the generated the question utterances if a human response was obtained from the microphone array. Then the next response utterance is generated from the results of this speech recognition by the open-domain chatbot "KACTUS". To improve the well-being and relationships of families or friends, The chatbot provides not only an enjoyable conversation directly related to the TV program, but also facilitates chatting with the robot regarding a wide range of related topics through a multiple dialog engine. The four following conversation engines were combined for response generation. 


\small
\begin{itemize}
\item {\verb|TV program|}: TV program conversation includes responses following a robot’s question utterance.
\item{\verb|Daily life|}: Broad conversation for generic content.
\item{\verb|News/SNS|}: Conversation using relevant internet news and Twitter comments that changes over time.
\item{\verb|Response generation|}: A transformer-based model~\cite{transformer001} that uses large-scale dialog-pair datasets.
\end{itemize}
\normalsize

\begin{figure}
  \includegraphics[width=0.8\textwidth]{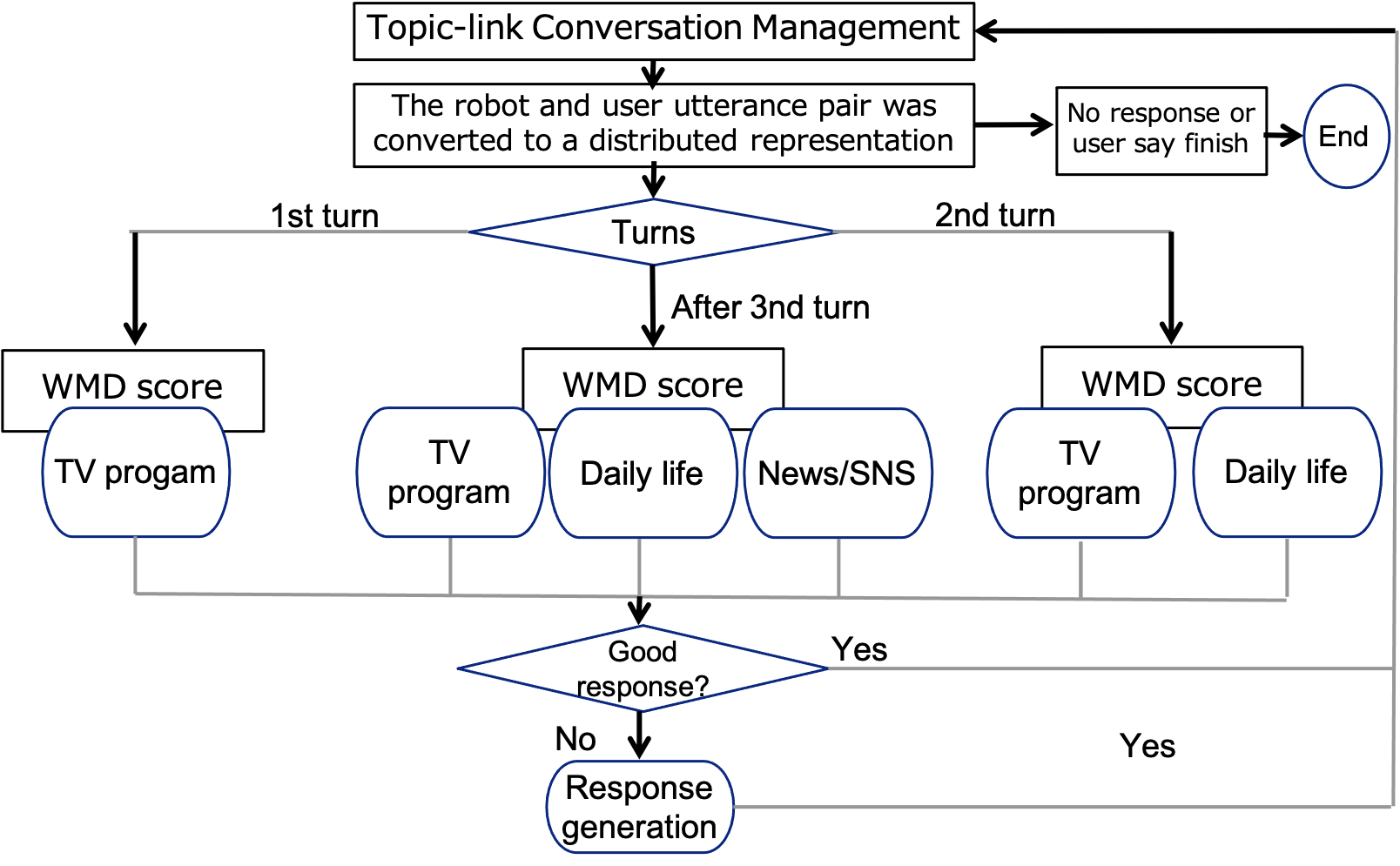}
  \caption{Function D: topic-linked conversation management.}
  \Description{Enjoying the baseball game from the third-base
  seats. Ichiro Suzuki preparing to bat.}
  \label{fig:flowchart}
\end{figure}

The chatbot strategy of topic-linked conversation management is shown in Figure 2 as follows.  
First, the robot and user utterance pair was converted to a distributed representation. Next, the robot selected the sentence with the highest Word Mover’s Distance~\cite{wmd001}, similar to the distributed representation from the different conversation engines, and used the sentence as an utterance. The 1st turn used only conversation engine for the TV program, the 2nd turn used the conversation engine for the TV program and Daily life, and after the third turn used the conversation engine for the TV program and Daily life and News/SNS. That is, as the conversation progresses, the topic is expected to gradually move away from the TV program to other wide topics such as daily life and the latest news/SNS. In these cases, the robot continued the conversation by switching to the three conversation engines mentioned above. 
When the Word Mover's Distance similarity between the user's utterance and the set of conversation engines falls below the threshold, the robot generates a response utterance using the transformer-based response generation engine to avoid interruption. Finally, the conversation would end if any of the following conditions are met, and then enter TV-watching mode.

\small
\begin{itemize}
\item {\verb||}If it is determined from the user's utterances that he or she wants to end the conversation.
\item {\verb||}If it is determined that the user has not answered more than twice.
\end{itemize}
\normalsize

When the conversation is continued, the robot waits once again for the voice input. When the conversation ends, the robot turned toward the TV and the current state transitions to the Status.1:TV watching.

\section{EXPERIMENTAL RESULTS}

To test user acceptance of the proposed conversation features designed for families or friends, we conduct the experiment in two steps. Although the same conversation functions of the robot are used, the first experiment focuses on "Research Question 1": can the robot's conversation features help promote communication between families or friends when they are watching TV? The second extended experiment then focuses on "Research Question 2": do the robot's conversation features improve the well-being and relationships of families or friends?


\subsection{Initial Experimental Results}

\subsubsection{Experimental setup} \hfill\\
The participants were assigned to one of three groups with two in individuals in each group, seen in the Table~\ref{tab:attributes}. We assume the initial use case is that the participants are enjoying the weekend TV programs, so the experiment time was approximately six hours from 10:00 to 16:00, including 1-hour lunch time. Participants were required to watch for the entire time, but can choose any TV program from the past seven years. At the end of the experiment, a 7-levels questionnaire survey ascending from "strongly disagree" to "strongly agree" in the Table~\ref{tab:results} and an interview were conducted. The experiments were conducted over two days for each group, the first day conduct with no robot and the second day conduct with robot.

In order to conduct the experiment in a relaxed state, we used a room for the experiment that mimicked a living room. The robot prototype was placed on a table (see Fig. 3 (a)) and worked independently without control by the experimenter. The disclosure and question utterance for the robot was set in a ratio of 3:1, and the utterance frequency was set to once every 80 seconds on average.

\begin{figure}
    \begin{subfigure}[h]{0.48\linewidth}
        \includegraphics[width=\linewidth]{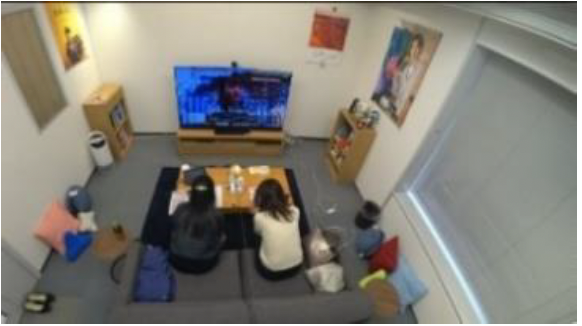}
        \vspace{-5mm}
        \caption{In the listening room}
    \end{subfigure}
    \hfill
    \begin{subfigure}[h]{0.48\linewidth}
        \includegraphics[width=\linewidth]{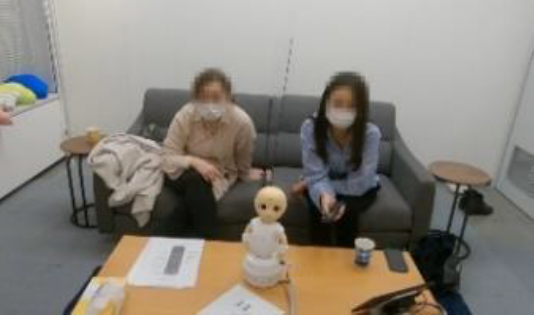}
        \vspace{-5mm}
        \caption{What it looks like in the experiment?}
    \end{subfigure}%
\vspace{-3mm}
\caption{Audiovisual experiment in progress}
\end{figure}

\begin{table}[!ht]
    \begin{minipage}[t]{0.48\linewidth}
      \caption{The attributes of participants}\label{tab:table1}
      \centering
      \footnotesize
      \vspace{-3.5mm}
      \begin{tabular}{|lllll|}
        \toprule
        \textbf{Participants}& \textbf{Gender,age}& \textbf{Relation} & \textbf{Talks}&\textbf{Robot}\\
        \midrule
        G1(no robot) & Female,30s & Family & 673 &-\\
        G1(with robot)    & Female,30s & Family & 555 &37\%\\
        G2(no robot) & Female,20s & Friends & 859 &-\\
        G2(with robot)    & Female,20s & Friends & 1446 &12\%\\
        G3(no robot) & Male,20s & Friends & 1337 &-\\
        G3(with robot)    & Male,20s & Friends & 960 &35\%\\
        \bottomrule
        \end{tabular}
        \label{tab:attributes}
    \end{minipage}
    \begin{minipage}[t]{0.48\linewidth}
      \caption{Results}\label{fig:table2}
      \centering
      \footnotesize
      \vspace{-3.5mm}
      \begin{tabular}{|clll|}
        \toprule
        \textbf{ID}& \textbf{Content}& \textbf{Score}& \textbf{SD}\\
        \midrule
        Q1 & \footnotesize Did you find conversational robots interesting? &6.5&0.46 \\
        Q2 &  \footnotesize What do you think of robot conversation? &5.0&1.51\\
        Q3-1 & Was the length of utterance appropriate? &5.0&1.77\\
        Q3-2 & Was the frequency of utterance appropriate? &3.5& 1.67\\
        Q3-3 & Was the timing of utterance appropriate? &3.5&1.75\\
        Q3-4 & Was the content of utterance appropriate? &2.5&1.16\\
        \bottomrule
        \end{tabular}
        \label{tab:results}
    \end{minipage}
\end{table}

\subsubsection{Analysis Results}
\paragraph{4.1.2.1  Number of human-utterances (talks)}
In Table~\ref{tab:attributes}, the number of human-utterances of G1 and G3 decreased with the robot, while the number increased with G2. The proportion of human-utterances triggered by the robot was 37\%, 12\%, and 35\%, respectively. Although we cannot conclude that the robot can increase communication opportunities, we found from the recorded videos that all groups listened to the robot's utterances with great interest and laughed much more frequently than without the robot. For example, when a person replied "I often shop at Ikebukuro", the robot responded with the news about a car accident at Ikebukuro (A famous commercial and entertainment district in Tokyo), whereupon two people wondered how the robot could have known about the car accident. In the future, We plan to use FER~\cite{yang001} quantitative analyses to show how does the robot change communication sentiment.

\paragraph{4.1.2.2  Questionnaire}
We design the questionnaire with the consideration of the users' interest and the acceptability of the utterances, by using 7-levels ascending from "strongly disagree" to "strongly agree". As for the users' interest, the Q1 and Q2 about overall experience and conversation features were rated 6.5 and 5.0 on average in Table~\ref{tab:results}, with standard deviations (SD) of 0.46 and 1.51, respectively. In the acceptance test of utterances in interviews, \textst{During the interview,} all participants said that the conversation with the robot was interesting enough. Then, the length, frequency, timing, and content of the robot's utterances were rated 5.0, 3.5, 3.5, and 2.5, respectively. In the interview, although most participants found the length of the utterances (Q3-1) to be "appropriate", they had a negative impression of the other three features, which was due to the fact that the robot was talking without considering that the participants were focusing on the TV program. They also claimed that the utterances generated sometimes seemed unnatural (e.g.,Do you eat beds?). Finally, we analyzed the conversation logs and found that the robot-human conversations consisted of 7.4 utterances in average, which is close to the average human-to-human conversations when watching TV~\cite{hoshi001}. Meanwhile, for the questionnaire item related to the content of the disclosure utterance(Q3-4), four out of six subjects responded negatively and a subject in group 2 said that “the robot sometimes said things which were unrelated to the TV program,” in the interview.
This also confirms how the effective the robot is in continuing a conversation. Considering these issues, we are developing an engagement AI method~\cite{wu001} so that the robot can register how intently users are watching TV. In addition, we plan to improve the accuracy of the utterance generation method in the future research.

\paragraph{4.1.2.3  The sample and the number of turns in the conversation between the subject and robot}

Finally, we verified the quality of the robot response utterance by "KACTUS" with reference to the number of turns in the conversation between the subject and robot. The number of utterances is given by the subject and robot for a continuous series of conversations regarding a single topic was defined as the number of turns and counted. Table~\ref{fig:questionnaire1} shows a continuous conversation sample and Table~\ref{fig:questionnaire2} shows the statistics of the turns. Human-to-human conversations consisted of approximately 3 to 6 turns, while on the other hand, 80\% of all the human-to-robot conversations consisted of approximately 5 to 8 turns. Especially group 3, which was presumed to have enjoyed the conversations with the robot, tended to have a higher number of turns in a single conversation compared to the other groups, and the robot’s effectiveness in continuing a conversation by switching the topic dialog was confirmed. 

\begin{table}[h!]
\caption{A conversation sample}
\vspace{-3.0mm}
\begin{tabular}{|c|l|c|}
\hline
\textbf{Speaker} & \textbf{Utterance content} & \textbf{dialog engine} \\ \hline
S: & Have you seen the Internet? & TV Program \\ \hline
U: &  Yes, I have! & - \\ \hline
S: & Oh, I see. Actually, I have, too. Wouldn't you like to see it sometime?  & TV Program\\ \hline
U: & Oh, I have seen it too? & -  \\ \hline
S: & Well, it is the Internet after all. I wonder what it's like? & Daily life \\ \hline
U: & It must be incredible, is not it, the Internet? & - \\ \hline
S: & \makecell[l]{According to CNET Japan, NVIDIA is close to an agreement to buy \\ Arm from Softbank."} & News/SNS \\ \hline
U: & Oh, really? What's that all about? & -  \\ \hline
S: & \makecell[l]{Softbank G wants to sell ARM to NVIDIA for 4.2 trillion yen. The deal \\is expected to close in 18 months.} & News/SNS \\ \hline
U: & That's a lot of time! & - \\ \hline
S: & TikTok will be a big hole Oracle acquisition. & News/SNS\\ \hline
U: & What about TikTok? & - \\ \hline
S: & NVIDIA announced an agreement with Softbank to buy ARM for \$40B. & News/SNS \\ \hline
U: & \$40B is cheap! & -\\ \hline
S: & I know, right? & Response generation\\ \hline
\end{tabular}
\label{fig:questionnaire1}
\end{table}

\begin{table}[H]
\caption{The conversation turns with robot}
\vspace{-3.0mm}
\begin{tabular}{|c|c|c|c|}
\hline
\multicolumn{4}{|c|}{The number of turns in the conversation between the subject and robot} \\ \hline
\textbf{Average}& 4.71& 5.26 & 6.88 \\ \hline
\textbf{Maximum}& 10& 11& 14\\ \hline
\end{tabular}
\label{fig:questionnaire2}
\end{table}


\paragraph{4.1.2.3  The conclusion and consideration of improvement for the extended experiment}

Although we cannot conclude from the result that the robot can increase communication between families and friends, the result shows that all participants enjoyed talking with the robot and the question about their interests in the robot were rated 6.5 (7-levels: ascending from "strongly disagree" to "strongly agree"). This indicates that watching TV with the robot has the potential to make our daily lives more enjoyable. On the other hand, considering the relatively low score from Q3-2 to Q3-4, we made three adjustments as follows for the next extended experiment.


 
\begin{itemize}
    \item We add a cancel button that allows the robot to stop talking immediately when the user is focused on a TV program such as a drama or movie. 
    \item We add more TV program words intto the dictionary for a better matching.
    \item The WMD threshold is set to a higher value to prevent the robot from talking about irrelevant topics of the TV program.
    
\end{itemize}

\subsection{Extended EXPERIMENTAL RESULTS}
To investigate the effect of the proposal on "Research Question 2": whether users feel that the robot's conversation features improve well-being and relationships among family or friends, we conduct the extended experiments as follows. We also revised the questionnaires so that we could assess the impact of TV watching robots on well-being and relationships. 

\subsubsection{Experimental Setup} \hfill\\
A total of 64 subjects, consisting of 32 groups of two individuals (families or friends), who were friendly with each other and preferred to watch TV. They were all recruited by a temporary employment agency, regardless of age or gender. The experiment outline is shown in Table~\ref{fig:questionnaire3} and the attributes of subjects are shown in Table~\ref{tab:QA}. As in the initial experiment, the experimental duration was approximately six hours including a one-hour lunch break, explanation and questionnaire time. The exact time allocation can be found in Table~\ref{fig:questionnaire3}. Subjects were allowed to watch any TV program they freely selected from the content server, which contains ten channels of recorded TV programs from the past seven years. A questionnaire survey and interview were conducted at the end of the experiment.

A randomized controlled experiment was conducted to investigate the effects of the presence or absence of a robot during TV watching. The intervention group was the group that watched TV with the robot, and the control group was the group that watched TV with humans only.

In order to conduct the experiment in a relaxed state, we used a room for the experiment that mimicked a living room. The robot prototype was placed on a table (see Fig. 3 (a)) and worked independently without control by the experimenter. The ratio of "TV-watching mode" and the "Conversation mode" for the robot was set to 3:1, and the frequency of utterances was set to once every 80 seconds on average. We use AWS to collect dialog data.

Before the experiment, we explained the purpose and procedure of the experiment. After the experiment, we conducted a questionnaire and interview to ask participants about their impressions of the experiment.

At the beginning of the experiment, the following instructions were given. Instruction text: "Today is a vacation. You are always busy, but today you decided to spend the day relaxing with your good friends (family), watching your favorite shows TV and chatting.


\begin{table}[H]
\caption{The outline of experiment}
\vspace{-3mm}
\begin{tabular}{|>{\centering\arraybackslash}m{1.5in}|l|}
\hline
\textbf{Subject}               &  \makecell[tl]{64 people (2 people a group, 32 groups) \\ Age: 10\textasciitilde 70s \\ Gender: male (9 people), female (55 people)} \\\hline
\textbf{Experiment period}     & \makecell[tl]{September\textasciitilde November, 2021 (Each group spent a day)} \\ \hline
\multirow{2}{*}{\textbf{Experiment time}} & \makecell[tl]{Morning (10:00-12:00) \\ 1) Explanation and confirmation of consent for the experiment consent form \\ 2) Watching TV without/with robot}  \\ \cline{2-2}  & \makecell[tl]{Afternoon (13: 00-18: 00) \\ 1) Watching TV without/with robot \\ 2) Post-questionnaire} \\ \hline
\end{tabular}
\end{table}

\subsubsection{Questionnaires} \hfill\\
The questionnaire that was asked as part of the viewing experiment consisted of the following two questions items, I-II. Questionnaires I were completed for the control groups, while questionnaire I-II was completed for the intervention groups after the experiment (Table. 6). The questions for questionnaire I-II are in Table. 7-8. 

\small
\begin{itemize}
\item{\verb|Questionnaire I|}: Question items to examine acceptance of general conversational robot impressions.
\item{\verb|Questionnaire II|}: Question items to examine the impressions of the proposed conversational features.
\end{itemize}
\normalsize

\begin{table}[H]
\caption{The conducted time and the conditions}
\vspace{-3mm}
\begin{tabular}{|l|l|l|}
\hline
 & \textbf{The control group} & \textbf{The intervention Group}\\ \hline
Morning (10:00-11:00) & \multicolumn{2}{|c|}{Explanation and confirmation of consent for the experiment consent form} \\ \hline
Morning (11:00-12:00) & Watching TV without robot & Watching TV without robot\\ \hline
Afternoon (13:00-17:00) & Watching TV without robot & Watching TV with robot\\ \hline
Afternoon (17:00-18:00) After the experiment& I & I, II \\ \hline
\end{tabular}
\label{fig:questionnaire3}
\end{table}


\begin{table}[htp]
  \begin{minipage}[htp]{.45\textwidth}
    \begin{center}
    \footnotesize
    \caption{Questionnaire I.}
    \vspace{-3mm}
    \begin{tabular}{|cl|}
    \toprule
    \textbf{No.} &  \textbf{Question} \\
    \midrule
    I-1  &  Would you like to have a conversation robot at home? \\ 
    \bottomrule
    \end{tabular}
    \end{center}
    \label{tab:subjects}
  \end{minipage}
  \hfill
  \begin{minipage}[htp]{.5\textwidth}
    \begin{center}
    \footnotesize
    \caption{Questionnaire II.}
    \vspace{-3mm}
    \begin{tabular}{|cl|}
    \toprule
    \textbf{No.} &  \textbf{Question} \\
    \midrule
    II-1  &  The conversational robot increased the communication between the two. \\ 
    II-2  &  The conversational robot made the place more relaxed. \\ 
    II-3  &  The conversational robot turned my attention to television. \\ 
    II-4  &  The conversational robot made it more interesting than usual. \\ 
    II-5  &  Do you think your daily life would be better if you had a conversational robot in your home\\
    II-6  &  How did you feel when the conversational robot was mumbling to itself?  \\ 
    II-7  &  How did you feel after interacting with the conversational robot?  \\ 
    II-8  &  Do you want the conversational robot at home if we make it better? \\
    II-9-1  &  Improvement expected: More and more talk related to the TV programs\\
    II-9-2  &  Improvement expected: The conversational robot can empathize with human emotions\\
    \bottomrule
    \end{tabular}
    \end{center}
    \label{tab:subjectsfeeling}
  \end{minipage}
\end{table}

\subsubsection{Analysis Results} \hfill\\
The following 2 factors were analysed based on the results of the questionnaire.
\small
\begin{itemize}
\item {\verb|Factor (1)|}: The acceptance of general conversational robots at home. 
\item{\verb|Factor (2)|}: The impact/impression of TV watching robots on well-being and relationships after the experiment
\end{itemize}
\normalsize
In analysis of factor(1), we compared the response results of questionnaire I in the control groups and in the intervention groups, and analysis of factor(2) is the response result of questionnaire II in the intervention groups.


\subsubsection{Analysis for Questionnaire I (Factor(1))} \hfill\\
In this study, we investigated which question items change the impression of general communication robots depending on whether they are used or not. Table 9. shows the results of the answers to the questionnaire I-1 "Would you like to have a talking robot at home?" (2-stage: 1: I think yes, 2: I do not think so), we found that the intervention group had a higher acceptance of the conversational robot with 43.75\% than the control group with 18.75\%. A chi-square test of I-1 was performed to detect significant differences between the two groups at a 5\% significance level. 
\small
\begin{itemize}
\item {\verb|(I-1)|}: Would you like to have a conversational robot at home?
\end{itemize}
\normalsize

\begin{table}[H]
\caption{The answer of I-1 "Would you like to have a talking robot at home?"}
\vspace{-3mm}
\begin{tabular}{|l|l|l|}
\hline
 & \textbf{The control group} & \textbf{The intervention Group}\\ \hline
I think so & 6 & 14\\ \hline
I don't think so & 26 & 18\\ \hline
Percentage of acceptance & 18.75\% & 43.75\% \\ \hline
\end{tabular}
\label{fig:questionnaire3}
\end{table}




\subsubsection{Analysis for Questionnaire II (Factor(2))} \hfill\\
To understand the user impressions and verify the effect of the designed conversational features A-D, we examined the responses of the questionnaire II given to the intervention groups in Table 10.

First, the answers to the questions II -1- II -5 in the questionnaire II are directly about whether the conversational features can improve well-being and relationships in the family or among friends. The response options were rated on a 7-point Likert scale from "7: very much" to "1: very little." We can see that II -1, II -2, and II -4 scored relatively high, while II -3 scored relatively low. From the additional comments, there are relatively many positive opinions for "the presence of the robot has strengthened the conversation between two people" and "the presence of the robot has relaxed the place". On the other hand, there are some negative opinions for "the robot's talking has drawn my attention to TV ". The current performance of the conversation functions did not have the effect of directing people's attention to TV while they were watching or interacting with other people.
As to the II-5: What do you think your daily life would be better if you had a robot in your home? About 75\% of the total respondents enjoyed interacting with the robot and. Most of them have positive comments included "I was able to enjoy watching TV shows with friends and family". 

Second, the answers to the questions II -6- II -7 . are directly about the user's impressions of the two modes of the conversational robot.

--II-6: How did you feel when the robot was talking to itself ("TV-watching mode")?
More than 72\% of the total respondents felt it was interesting. There were also some negative comments such as "sometimes it did not match the program" and "sometimes it said the same thing again".

--II-7: How did you feel after interacting with the robot ("Conversation mode")?
About 76\% of the total respondents enjoyed interacting with the robot. Positive comments included "they enjoyed the conversation between the robot and their family or friends". Negative comments included "it did not match the TV program sometimes" and "sometimes it said the same topic repeatedly".
As described above, about 70\% of respondents thought that the robot's conversation were interesting, but there were some problems, such as content overlap.

Third, the answers to the questions provide information about user acceptance and the improvement factor in the future. 
-- II -8 shows a higher acceptance rate than the result 43.75\% in I-1 with 72\% that users would like to have the conversational robot if we improved the conversational features.
-- II -9-1- II -9-2 shows the users' expected improvement factor of top-2 for the conversation functions in the future.


\begin{table}[h!]\footnotesize
\caption{The answer results of questionnaire questions II-1 to II-9}
\vspace{-3.0mm}
\setlength\tabcolsep{3.8pt}
\begin{tabular}{|c|c|c|c|}
\hline
\textbf{Questionnaire Questions} & \textbf{Percentage of "Yes"} & \textbf{Average of score} & \textbf{SD} \\ \hline 
\makecell[tl]{II-1: The conversational robot increased the communication between the two?} 
&- & 4.78 & 1.38 \\ \hline
\makecell[tl]{II-2: The conversational robot made the place more relaxed.}
&- & 5.06 & 1.29 \\ \hline
\makecell[tl]{II-3: The conversational robot turned my attention to television.}
&- & 3.16 & 1.43 \\ \hline
\makecell[tl]{II-4: The conversational robot made it more interesting than usual.}
&- & 4.5 & 1.39 \\ \hline
\makecell[tl]{II-5: Do you think your daily life would be better if you had a conversational robot in your home?}
&75\% &-&-\\ \hline

\makecell[tl]{II-6: How did you feel when the conversational robot was mumbling to itself? }
&72\% &-&-\\ \hline
\makecell[tl]{II-7: How did you feel when you interacted with the conversational robot?}
&76\% &-&-\\ \hline

\makecell[tl]{II-8: Do you want the conversational robot at home if we make it better? }
&72\% &-&-\\ \hline

\makecell[tl]{II-9-1: Improvement expected: More and more talk related to the TV programs}
&- & 5.03 & 1.23 \\ \hline
\makecell[tl]{II-9-2: Improvement expected: The conversational robot can know/empathize with human emotions}
&- & 5.22 & 1.61 \\ \hline

\end{tabular}
\end{table}

Finally Table~\ref{tab:statistics}. shows the Analysis for Conversations turns.

\begin{table}[h!]
\caption{The statistics of the conversation turns }
\vspace{-3.0mm}
\footnotesize
\setlength\tabcolsep{3.8pt} 
\begin{tabular}{|c|c|c|c|c|c|c|c|c|c|c|c|c|c|c|c|c|c|c|c|}
\hline
\textbf{Turns}&Total&9/1&9/8&9/14&9/18&9/23&9/30&10/5&10/13&10/16&10/19&10/20&10/23&10/27&10/28&11/9&11/12&11/18&11/19 \\ \hline 
\textbf{Mean}& 5.54& 	4.92& 	4.15& 	7.00& 	6.32& 	5.96& 	5.45& 	5.38& 	6.45& 	5.12& 	5.67& 	6.83& 	5.91& 	5.40& 	6.21& 	5.88& 	6.52  & 	4.09 & 4.61  \\ \hline
\textbf{Max}& 16&	8& 	8& 	10& 	11& 	8& 	10& 	9& 	10& 	12& 	8& 	13& 	16& 	9& 	9& 	9& 	11 &7&9 \\ \hline
\end{tabular}
\label{tab:statistics}
\end{table}

As in the initial experiment, Table~\ref{tab:statistics} shows the statistics of the turns in the extended experiment.
We found that the average and maximum number of turns in the conversation did not change significantly, also indicating that the intervention groups enjoyed the conversations with the robot and that the robot's effectiveness in continuing a conversation was confirmed by the topic-linked conversation management method.

\subsubsection{Consideration} \hfill\\
From the result, the intervention groups showed higher demand for "It would be nice to have a TV-watching robot at home" compared with the control groups. The result also indicates that proposed conversational features have the potential to improve the well-being and relationships of families or friends. Furthermore, we also obtain improvement points in the future which will make a high user acceptance to the conversational robot, such as the conversation timing when the user focuses on the TV, conversation topics that are not consistent with the TV program or unsuitable topic transitions. We will improve the robot to resolve these issues and conduct future experiments with more subjects to verify the effect of our robot on humans while watching TV.

\section{CONCLUSION}
In this study, we presented the functional requirements of a robot that watches TV with people based on the dialog analyses between people when watching TV. In addition, we assessed the robot prototype developed as the first step in realizing this robot based on these functional requirements and the results of the operational verification based on TV watching experiments. Inspired by conversational analysis of personal conversation behaviors when watching TV, the robot works by switching between the two following modes, "TV-watching mode" and "Conversation mode”. In the "TV-watching mode.", the robot first extracts keywords from the TV program, and then generates the disclosure utterances based on the extracted keywords as if enjoying the TV program. In the "conversation mode", the robot generates question utterances with keywords in the same way, and then employs a topics-based dialog management method consisting of multiple dialog engines for rich conversations related to the TV-program. 

The initial experimental were conducted in which all participants are divided into three groups with two participants in each group. Although we cannot conclude that the robot can increase communication opportunities between families or friends when they are watching TV, the result shows that all participants from the three groups enjoyed talking with the robot. This also indicates that the proposed conversational features of TV-watching Companion Robot have the potential to make our daily lives more enjoyable. 

Furthermore, we further conduct additional extended experiments with 32 groups with two participants in each group, half of whom watched TV without the robot as the control group and the other half watches TV with the robot as the intervention groups. The intervention groups showed higher demand for "It would be nice to have a TV-watching robot at home" compared with the control groups, and a significant difference was confirmed. We also found that many of the intervention groups felt that the conversational functions of the robot "made the place more relaxed" and "increased conversation between the two of us," and "made it more interesting than usual.". The result indicates that proposed conversational features have the potential to improve the well-being and relationships of families or friends. This study also obtain improvement points based on the questionnaire and operational review.

Meanwhile, we collected points for improvements based on the operational review, such as the conversation timing when the user focuses on the TV, conversation topics that are not consistent with the TV program or unsuitable topic transitions. We will improve the robot to resolve these issues and conduct future experiments with more subjects to verify the effect of our robot on humans while watching TV.

\begin{acks}
The authors would like to thank to the experts from Systemsoft Inc: Zhiguang Zhou, Zheshuang Lyu, Wataru Nishioka, Megumi Komiya for their great support and advice on our proposed method and its optimization. 
\end{acks}

\bibliographystyle{ACM-Reference-Format}
\bibliography{sample-base}


\begin{thebibliography}{33}


\ifx \showCODEN    \undefined \def \showCODEN     #1{\unskip}     \fi
\ifx \showDOI      \undefined \def \showDOI       #1{#1}\fi
\ifx \showISBNx    \undefined \def \showISBNx     #1{\unskip}     \fi
\ifx \showISBNxiii \undefined \def \showISBNxiii  #1{\unskip}     \fi
\ifx \showISSN     \undefined \def \showISSN      #1{\unskip}     \fi
\ifx \showLCCN     \undefined \def \showLCCN      #1{\unskip}     \fi
\ifx \shownote     \undefined \def \shownote      #1{#1}          \fi
\ifx \showarticletitle \undefined \def \showarticletitle #1{#1}   \fi
\ifx \showURL      \undefined \def \showURL       {\relax}        \fi
\providecommand\bibfield[2]{#2}
\providecommand\bibinfo[2]{#2}
\providecommand\natexlab[1]{#1}
\providecommand\showeprint[2][]{arXiv:#2}

\bibitem[Azure(2021)]%
        {azure001}
\bibfield{author}{\bibinfo{person}{Microsoft Azure}.}
  \bibinfo{year}{2021}\natexlab{}.
\newblock \bibinfo{booktitle}{\emph{Azure Cognitive Services}}.
\newblock Microsoft.
\newblock
\urldef\tempurl%
\url{https://azure.microsoft.com/en-us/services/cognitive-services/}
\showURL{%
Retrieved Oct 3, 2021 from \tempurl}


\bibitem[Hagio et~al\mbox{.}(2021)]%
        {hagio001}
\bibfield{author}{\bibinfo{person}{Yuta Hagio}, \bibinfo{person}{Marina
  Kamimura}, \bibinfo{person}{Yuta Hoshi}, \bibinfo{person}{Yutaka Kaneko},
  {and} \bibinfo{person}{Masao Yamamoto}.} \bibinfo{year}{2021}\natexlab{}.
\newblock \showarticletitle{TV-watching Robot: Toward Enriching Media
  Experience and Activating Human Communication}. In
  \bibinfo{booktitle}{\emph{International Broadcasting Convention, 2021}}.
\newblock


\bibitem[Hidekazu~Minami and Hagita(2016)]%
        {minami001}
\bibfield{author}{\bibinfo{person}{Masayuki~Kanbara Hidekazu~Minami,
  Hiromichi~Kawanami} {and} \bibinfo{person}{Norihiro Hagita}.}
  \bibinfo{year}{2016}\natexlab{}.
\newblock \showarticletitle{Chat robot coupling machine responses and social
  media comments for continuous conversation}. In \bibinfo{booktitle}{\emph{In
  2016 IEEE International Conference on Multimedia Expo Workshops (ICMEW).}}
  \bibinfo{publisher}{IEEE Press}, \bibinfo{pages}{1--6}.
\newblock


\bibitem[Hoshi et~al\mbox{.}(2020)]%
        {hoshi001}
\bibfield{author}{\bibinfo{person}{Yuta Hoshi}, \bibinfo{person}{Yutaka
  Kaneko}, \bibinfo{person}{Michihiro Uehara}, \bibinfo{person}{Yuta Hagio},
  \bibinfo{person}{Yasuhiro Murasaki}, \bibinfo{person}{Satoshi Nishimura},
  {and} \bibinfo{person}{Masao Yamamoto}.} \bibinfo{year}{2020}\natexlab{}.
\newblock \showarticletitle{Utterance Function for Companion Robot for Humans
  Watching Television}. In \bibinfo{booktitle}{\emph{2020 {IEEE} International
  Conference on Consumer Electronics (ICCE), Las Vegas, NV, USA, January 4-6,
  2020}}. \bibinfo{publisher}{IEEE}, \bibinfo{address}{Las Vegas, NV, USA},
  \bibinfo{pages}{1--5}.
\newblock


\bibitem[House et~al\mbox{.}(1988)]%
        {house1988social}
\bibfield{author}{\bibinfo{person}{James~S House}, \bibinfo{person}{Karl~R
  Landis}, {and} \bibinfo{person}{Debra Umberson}.}
  \bibinfo{year}{1988}\natexlab{}.
\newblock \showarticletitle{Social relationships and health}.
\newblock \bibinfo{journal}{\emph{Science}} \bibinfo{volume}{241},
  \bibinfo{number}{4865} (\bibinfo{year}{1988}), \bibinfo{pages}{540--545}.
\newblock


\bibitem[Iio et~al\mbox{.}(2020)]%
        {iio2020twin}
\bibfield{author}{\bibinfo{person}{Takamasa Iio}, \bibinfo{person}{Yuichiro
  Yoshikawa}, \bibinfo{person}{Mariko Chiba}, \bibinfo{person}{Taichi Asami},
  \bibinfo{person}{Yoshinori Isoda}, {and} \bibinfo{person}{Hiroshi Ishiguro}.}
  \bibinfo{year}{2020}\natexlab{}.
\newblock \showarticletitle{Twin-robot dialogue system with robustness against
  speech recognition failure in human-robot dialogue with elderly people}.
\newblock \bibinfo{journal}{\emph{Applied Sciences}} \bibinfo{volume}{10},
  \bibinfo{number}{4} (\bibinfo{year}{2020}), \bibinfo{pages}{1522}.
\newblock


\bibitem[Jokinen et~al\mbox{.}(2019)]%
        {jokinen2019human}
\bibfield{author}{\bibinfo{person}{Kristiina Jokinen}, \bibinfo{person}{Satoshi
  Nishimura}, \bibinfo{person}{Kentaro Watanabe}, {and}
  \bibinfo{person}{Takuichi Nishimura}.} \bibinfo{year}{2019}\natexlab{}.
\newblock \showarticletitle{Human-robot dialogues for explaining activities}.
  In \bibinfo{booktitle}{\emph{9th International Workshop on Spoken Dialogue
  System Technology}}. Springer, \bibinfo{pages}{239--251}.
\newblock


\bibitem[Kidd and Breazeal(2004)]%
        {kidd001}
\bibfield{author}{\bibinfo{person}{Cory~D. Kidd} {and} \bibinfo{person}{Cynthia
  Breazeal}.} \bibinfo{year}{2004}\natexlab{}.
\newblock \showarticletitle{Effect of a robot on user perceptions}. In
  \bibinfo{booktitle}{\emph{2004 {IEEE/RSJ} International Conference on
  Intelligent Robots and Systems, Sendai, Japan, September 28 - October 2,
  2004}}. \bibinfo{publisher}{{IEEE}}, \bibinfo{address}{Sendai, Japan},
  \bibinfo{pages}{3559--3564}.
\newblock


\bibitem[Kudo(2006)]%
        {mecab001}
\bibfield{author}{\bibinfo{person}{Taku Kudo}.}
  \bibinfo{year}{2006}\natexlab{}.
\newblock \bibinfo{booktitle}{\emph{MeCab: Yet Another Part-of-speech and
  Morphological Analyzer}}.
\newblock NAIST.
\newblock
\urldef\tempurl%
\url{http://taku910.github.io/mecab/}
\showURL{%
Retrieved Oct 3, 2021 from \tempurl}


\bibitem[Kusner et~al\mbox{.}(2015)]%
        {wmd001}
\bibfield{author}{\bibinfo{person}{Matt~J. Kusner}, \bibinfo{person}{Yu Sun},
  \bibinfo{person}{Nicholas~I. Kolkin}, {and} \bibinfo{person}{Kilian~Q.
  Weinberger}.} \bibinfo{year}{2015}\natexlab{}.
\newblock \showarticletitle{From Word Embeddings To Document Distances}. In
  \bibinfo{booktitle}{\emph{Proceedings of the 32nd International Conference on
  Machine Learning, {ICML} 6-11}} \emph{(\bibinfo{series}{{JMLR} Workshop and
  Conference Proceedings}, Vol.~\bibinfo{volume}{37})}.
  \bibinfo{publisher}{JMLR.org}, \bibinfo{address}{Lille, France},
  \bibinfo{pages}{957--966}.
\newblock


\bibitem[Lin et~al\mbox{.}(2014a)]%
        {coco001r}
\bibfield{author}{\bibinfo{person}{Tsung{-}Yi Lin}, \bibinfo{person}{Michael
  Maire}, \bibinfo{person}{Serge~J. Belongie}, \bibinfo{person}{James Hays},
  \bibinfo{person}{Pietro Perona}, \bibinfo{person}{Deva Ramanan},
  \bibinfo{person}{Piotr Doll{\'{a}}r}, {and} \bibinfo{person}{C.~Lawrence
  Zitnick}.} \bibinfo{year}{2014}\natexlab{a}.
\newblock \showarticletitle{Microsoft {COCO:} Common Objects in Context}. In
  \bibinfo{booktitle}{\emph{Computer Vision - {ECCV} 2014 - 13th European
  Conference, Zurich, Switzerland, September 6-12, 2014, Proceedings, Part
  {V}}} \emph{(\bibinfo{series}{Lecture Notes in Computer Science},
  Vol.~\bibinfo{volume}{8693})}, \bibfield{editor}{\bibinfo{person}{David~J.
  Fleet}, \bibinfo{person}{Tom{\'{a}}s Pajdla}, \bibinfo{person}{Bernt
  Schiele}, {and} \bibinfo{person}{Tinne Tuytelaars}} (Eds.).
  \bibinfo{publisher}{Springer}, \bibinfo{address}{Zurich, Switzerland},
  \bibinfo{pages}{740--755}.
\newblock
\urldef\tempurl%
\url{https://doi.org/10.1007/978-3-319-10602-1\_48}
\showDOI{\tempurl}


\bibitem[Lin et~al\mbox{.}(2014b)]%
        {coco001}
\bibfield{author}{\bibinfo{person}{Tsung-Yi Lin}, \bibinfo{person}{Michael
  Maire}, \bibinfo{person}{Serge Belongie}, \bibinfo{person}{James Hays},
  \bibinfo{person}{Pietro Perona}, \bibinfo{person}{Deva Ramanan},
  \bibinfo{person}{Piotr Doll{\'a}r}, {and} \bibinfo{person}{C~Lawrence
  Zitnick}.} \bibinfo{year}{2014}\natexlab{b}.
\newblock \showarticletitle{Microsoft coco: Common objects in context}. In
  \bibinfo{booktitle}{\emph{European conference on computer vision}}.
  \bibinfo{publisher}{Springer}, \bibinfo{address}{Zurich, Switzerland},
  \bibinfo{pages}{740--755}.
\newblock


\bibitem[Mathis et~al\mbox{.}(2020)]%
        {ARTV002}
\bibfield{author}{\bibinfo{person}{Florian Mathis}, \bibinfo{person}{Xuesong
  Zhang}, \bibinfo{person}{Mark McGill}, \bibinfo{person}{Adalberto Simeone},
  {and} \bibinfo{person}{Mohamed Khamis}.} \bibinfo{year}{2020}\natexlab{}.
\newblock \showarticletitle{Assessing Social Text Placement in Mixed Reality
  TV}. In \bibinfo{booktitle}{\emph{ACM International Conference on Interactive
  Media Experiences (IMX'20)}}. \bibinfo{publisher}{ACM},
  \bibinfo{address}{Barcelona, Spain}, \bibinfo{pages}{205--211}.
\newblock


\bibitem[Mikolov et~al\mbox{.}(2013)]%
        {vec001}
\bibfield{author}{\bibinfo{person}{Tomas Mikolov}, \bibinfo{person}{Ilya
  Sutskever}, \bibinfo{person}{Kai Chen}, \bibinfo{person}{Greg~S Corrado},
  {and} \bibinfo{person}{Jeff Dean}.} \bibinfo{year}{2013}\natexlab{}.
\newblock \showarticletitle{Distributed representations of words and phrases
  and their compositionality}. In \bibinfo{booktitle}{\emph{Advances in neural
  information processing systems}}. \bibinfo{publisher}{In Proceedings of the
  26th International Conference on Neural Information Processing Systems},
  \bibinfo{address}{Lake Tahoe, Nevada, United States},
  \bibinfo{pages}{3111--3119}.
\newblock


\bibitem[Minami et~al\mbox{.}(2016)]%
        {minami2016chat}
\bibfield{author}{\bibinfo{person}{Hidekazu Minami}, \bibinfo{person}{Hiromichi
  Kawanami}, \bibinfo{person}{Masayuki Kanbara}, {and}
  \bibinfo{person}{Norihiro Hagita}.} \bibinfo{year}{2016}\natexlab{}.
\newblock \showarticletitle{Chat robot coupling machine responses and social
  media comments for continuous conversation}. In
  \bibinfo{booktitle}{\emph{2016 IEEE International Conference on Multimedia \&
  Expo Workshops (ICMEW)}}. IEEE, \bibinfo{pages}{1--6}.
\newblock


\bibitem[Muto et~al\mbox{.}(2006)]%
        {muto001}
\bibfield{author}{\bibinfo{person}{Yuki Muto}, \bibinfo{person}{Yoshihiro
  Iwase}, \bibinfo{person}{Shunichi Hattori}, \bibinfo{person}{Yasufumi
  Takama}, {and} \bibinfo{person}{Kaoru Hirota}.}
  \bibinfo{year}{2006}\natexlab{}.
\newblock \showarticletitle{Web intelligence approach for human robot
  communication under TV watching environment}. In
  \bibinfo{booktitle}{\emph{SCIS \& ISIS SCIS \& ISIS 2006}}. Japan Society for
  Fuzzy Theory and Intelligent Informatics, \bibinfo{publisher}{ACM},
  \bibinfo{address}{Japan}, \bibinfo{pages}{426--429}.
\newblock


\bibitem[Nakadai et~al\mbox{.}(2017)]%
        {hark001}
\bibfield{author}{\bibinfo{person}{Kazuhiro Nakadai},
  \bibinfo{person}{Hiroshi~G. Okuno}, {and} \bibinfo{person}{Takeshi
  Mizumoto.}} \bibinfo{year}{2017}\natexlab{}.
\newblock \showarticletitle{Development, Deployment and Applications of Robot
  Audition Open Source Software HARK}.
\newblock \bibinfo{journal}{\emph{Journal of Robotics and Mechatronics}}
  \bibinfo{volume}{29}, \bibinfo{number}{1} (\bibinfo{year}{2017}),
  \bibinfo{pages}{16--25}.
\newblock


\bibitem[Ogawa et~al\mbox{.}(2018)]%
        {ogawa001}
\bibfield{author}{\bibinfo{person}{Hiromu Ogawa}, \bibinfo{person}{Masaya
  Ikeo}, \bibinfo{person}{Hisayuki Ohmata}, \bibinfo{person}{Chigusa Yamamura},
  {and} \bibinfo{person}{Hiroshi Fujisawa}.} \bibinfo{year}{2018}\natexlab{}.
\newblock \showarticletitle{System architecture for IoT services with broadcast
  content}. In \bibinfo{booktitle}{\emph{{IEEE} International Conference on
  Consumer Electronics, {ICCE} 2018, Las Vegas, NV, USA, January 12-14, 2018}}.
  \bibinfo{publisher}{{IEEE}}, \bibinfo{address}{Las Vegas, NV},
  \bibinfo{pages}{1--2}.
\newblock


\bibitem[on~the Ageing Society: 2020~(Summary)<FY2020>(2020)]%
        {Annualreport}
\bibfield{author}{\bibinfo{person}{Japan Cabinet Office: Annual~Report on~the
  Ageing Society: 2020~(Summary)<FY2020>}.} \bibinfo{year}{2020}\natexlab{}.
\newblock \bibinfo{title}{Establishing a New Paradigm of Health through Science
  and Technology in Ageing Society.}
\newblock
\newblock
\urldef\tempurl%
\url{https://www8.cao.go.jp/kourei/english/annualreport/2020/pdf/2020.pdf
  (accessed on 23 January 2022)}
\showURL{%
\tempurl}


\bibitem[Pollack et~al\mbox{.}(2002)]%
        {pollack2002pearl}
\bibfield{author}{\bibinfo{person}{Martha~E Pollack}, \bibinfo{person}{Laura
  Brown}, \bibinfo{person}{Dirk Colbry}, \bibinfo{person}{Cheryl Orosz},
  \bibinfo{person}{Bart Peintner}, \bibinfo{person}{Sailesh Ramakrishnan},
  \bibinfo{person}{Sandra Engberg}, \bibinfo{person}{Judith~T Matthews},
  \bibinfo{person}{Jacqueline Dunbar-Jacob}, \bibinfo{person}{Colleen~E
  McCarthy}, {et~al\mbox{.}}} \bibinfo{year}{2002}\natexlab{}.
\newblock \showarticletitle{Pearl: A mobile robotic assistant for the elderly}.
  In \bibinfo{booktitle}{\emph{AAAI workshop on automation as eldercare}},
  Vol.~\bibinfo{volume}{2002}. \bibinfo{pages}{85--91}.
\newblock


\bibitem[Saeghe et~al\mbox{.}(2019)]%
        {ARTV001}
\bibfield{author}{\bibinfo{person}{Pejman Saeghe}, \bibinfo{person}{Sarah
  Clinc}, \bibinfo{person}{Bruce Weir}, \bibinfo{person}{Maxine Glancy},
  \bibinfo{person}{Vinoba Vinayagamoorthy}, \bibinfo{person}{Ollie Pattinson},
  \bibinfo{person}{Stephen Pettifer}, {and} \bibinfo{person}{Robert Stevens}.}
  \bibinfo{year}{2019}\natexlab{}.
\newblock \showarticletitle{Augmenting Television With Augmented Reality}. In
  \bibinfo{booktitle}{\emph{2019 ACM International Conference on Interactive
  Experiences for TV and Online Video (TVX'19)}}. \bibinfo{publisher}{ACM},
  \bibinfo{address}{Salford (Manchester), UK}, \bibinfo{pages}{255--261}.
\newblock


\bibitem[Sato et~al\mbox{.}(2017)]%
        {mecab002}
\bibfield{author}{\bibinfo{person}{Toshinori Sato}, \bibinfo{person}{Taiichi
  Hashimoto}, {and} \bibinfo{person}{Manabu Okumura}.}
  \bibinfo{year}{2017}\natexlab{}.
\newblock \showarticletitle{Implementation of a word segmentation dictionary
  called mecab-ipadic-NEologd and study on how to use it effectively for
  information retrieval}. In \bibinfo{booktitle}{\emph{Proceedings of the
  twenty-three annual meeting of the association for natural language
  processing}}. The Association for Natural Language Processing,
  \bibinfo{publisher}{In Annual Meeting of the Association for Natural Language
  Processing.}, \bibinfo{address}{Japan}, \bibinfo{pages}{NLP2017--B6}.
\newblock


\bibitem[Service(2021)]%
        {aws001}
\bibfield{author}{\bibinfo{person}{Amazon~Web Service}.}
  \bibinfo{year}{2021}\natexlab{}.
\newblock \bibinfo{booktitle}{}.
\newblock Amazon.
\newblock
\urldef\tempurl%
\url{https://aws.amazon.com/}
\showURL{%
Retrieved Oct 3, 2021 from \tempurl}


\bibitem[Shaoqing et~al\mbox{.}(2017)]%
        {rcnn001}
\bibfield{author}{\bibinfo{person}{Ren Shaoqing}, \bibinfo{person}{He Kaiming},
  \bibinfo{person}{Girshick Ross}, {and} \bibinfo{person}{Sun Jian}.}
  \bibinfo{year}{2017}\natexlab{}.
\newblock \showarticletitle{Faster R-CNN: Towards Real-Time Object Detection
  with Region Proposal Networks}.
\newblock \bibinfo{journal}{\emph{IEEE Transactions on Pattern Analysis and
  Machine Intelligence}} \bibinfo{volume}{39}, \bibinfo{number}{6}
  (\bibinfo{year}{2017}), \bibinfo{pages}{1137--1149}.
\newblock


\bibitem[Shogo~Nishimura and Hagita(2017)]%
        {nishimura001}
\bibfield{author}{\bibinfo{person}{Masayuki~Kanbara Shogo~Nishimura,
  Hiromichi~Kawanami} {and} \bibinfo{person}{Norihiro Hagita}.}
  \bibinfo{year}{2017}\natexlab{}.
\newblock \showarticletitle{A TV Chat Robot with Time-Shifting Function for
  Daily-Use Communication}. In \bibinfo{booktitle}{\emph{In The 9th
  International Conference on Social Robotics (ICSR).}}
  \bibinfo{publisher}{Springer}, \bibinfo{address}{Tsukuba, Japan},
  \bibinfo{pages}{516--525}.
\newblock


\bibitem[Sok et~al\mbox{.}(2019)]%
        {sok001}
\bibfield{author}{\bibinfo{person}{Sohyune~R Sok}, \bibinfo{person}{Mi~Hyeon
  Seong}, {and} \bibinfo{person}{Mi~Hye Ryu}.} \bibinfo{year}{2019}\natexlab{}.
\newblock \showarticletitle{Differences of self-control, daily life stress, and
  communication skills between smartphone addiction risk group and general
  group in Korean nursing students}.
\newblock \bibinfo{journal}{\emph{Psychiatric Quarterly}} \bibinfo{volume}{90},
  \bibinfo{number}{1} (\bibinfo{year}{2019}), \bibinfo{pages}{1--9}.
\newblock


\bibitem[Su et~al\mbox{.}(2017)]%
        {su2017chatbot}
\bibfield{author}{\bibinfo{person}{Ming-Hsiang Su},
  \bibinfo{person}{Chung-Hsien Wu}, \bibinfo{person}{Kun-Yi Huang},
  \bibinfo{person}{Qian-Bei Hong}, {and} \bibinfo{person}{Hsin-Min Wang}.}
  \bibinfo{year}{2017}\natexlab{}.
\newblock \showarticletitle{A chatbot using LSTM-based multi-layer embedding
  for elderly care}. In \bibinfo{booktitle}{\emph{2017 International Conference
  on Orange Technologies (ICOT)}}. IEEE, \bibinfo{pages}{70--74}.
\newblock


\bibitem[University(2015)]%
        {commu001}
\bibfield{author}{\bibinfo{person}{Osaka University}.}
  \bibinfo{year}{2015}\natexlab{}.
\newblock \bibinfo{booktitle}{\emph{Development of Social Dialogue Robots
  “CommU” and “Sota”}}.
\newblock Osaka University.
\newblock
\urldef\tempurl%
\url{https://resou.osaka-u.ac.jp/en/research/2015/20150120_2}
\showURL{%
Retrieved Oct 3, 2021 from \tempurl}


\bibitem[Vaswani et~al\mbox{.}(2017)]%
        {transformer001}
\bibfield{author}{\bibinfo{person}{Ashish Vaswani}, \bibinfo{person}{Noam
  Shazeer}, \bibinfo{person}{Niki Parmar}, \bibinfo{person}{Jakob Uszkoreit},
  \bibinfo{person}{Llion Jones}, \bibinfo{person}{Aidan~N Gomez},
  \bibinfo{person}{{\L}ukasz Kaiser}, {and} \bibinfo{person}{Illia
  Polosukhin}.} \bibinfo{year}{2017}\natexlab{}.
\newblock \showarticletitle{Attention is all you need}. In
  \bibinfo{booktitle}{\emph{Advances in neural information processing
  systems}}. \bibinfo{publisher}{Curran Associates Inc.},
  \bibinfo{address}{Long Beach, CA}, \bibinfo{pages}{5998--6008}.
\newblock


\bibitem[Wu et~al\mbox{.}(2020a)]%
        {kactus001}
\bibfield{author}{\bibinfo{person}{Jianming Wu}, \bibinfo{person}{Bo Yang},
  \bibinfo{person}{Gen Hattori}, {and} \bibinfo{person}{Yasuhiro Takishima}.}
  \bibinfo{year}{2020}\natexlab{a}.
\newblock \bibinfo{booktitle}{\emph{Chatbot AI 'KACTUS' by Utilizing Multimodal
  Recognition Technologies}}.
\newblock \bibinfo{type}{SIG Technical Reports, AVM} 2021(7), 1-5.
  \bibinfo{institution}{Information Processing Society of Japan}.
\newblock


\bibitem[Wu et~al\mbox{.}(2020b)]%
        {wu001}
\bibfield{author}{\bibinfo{person}{Jianming Wu}, \bibinfo{person}{Bo Yang},
  \bibinfo{person}{Yanan Wang}, {and} \bibinfo{person}{Gen Hattori}.}
  \bibinfo{year}{2020}\natexlab{b}.
\newblock \showarticletitle{Advanced Multi-Instance Learning Method with
  Multi-features Engineeringand Conservative Optimization for Engagement
  Intensity Prediction}. In \bibinfo{booktitle}{\emph{{ICMI} '20: International
  Conference on Multimodal Interaction, Virtual Event, The Netherlands, October
  25-29, 2020}}. \bibinfo{publisher}{{ACM}}, \bibinfo{address}{The
  Netherlands}, \bibinfo{pages}{777--783}.
\newblock


\bibitem[Yamamoto et~al\mbox{.}(2009)]%
        {yamamoto2009dialogue}
\bibfield{author}{\bibinfo{person}{Daisuke Yamamoto}, \bibinfo{person}{Yuka
  Kobayashi}, \bibinfo{person}{Sachie Yokoyama}, {and} \bibinfo{person}{Miwako
  Doi}.} \bibinfo{year}{2009}\natexlab{}.
\newblock \showarticletitle{Dialogue Interface for Elderly People--Robot
  becomes a partner and makes the elders happy}.
\newblock \bibinfo{journal}{\emph{IEICE Technical Report; IEICE Tech. Rep.}}
  \bibinfo{volume}{109}, \bibinfo{number}{224} (\bibinfo{year}{2009}),
  \bibinfo{pages}{47--51}.
\newblock


\bibitem[Yang et~al\mbox{.}(2021)]%
        {yang001}
\bibfield{author}{\bibinfo{person}{Bo Yang}, \bibinfo{person}{Wu Jianming},
  {and} \bibinfo{person}{Gen Hattori}.} \bibinfo{year}{2021}\natexlab{}.
\newblock \showarticletitle{Face Mask Aware Robust Facial Expression
  Recognition During The Covid-19 Pandemic}. In \bibinfo{booktitle}{\emph{2021
  IEEE International Conference on Image Processing (ICIP)}}.
  \bibinfo{publisher}{IEEE}, \bibinfo{address}{Anchorage, Alaska, USA},
  \bibinfo{pages}{240--244}.
\newblock


\end{thebibliography}

\appendix

\end{document}